\documentclass[sigconf]{acmart}

\renewcommand\footnotetextcopyrightpermission[1]{
\footnotetext{Permission to make digital or hard copies of part or all of this work
   for personal or classroom use is granted without fee provided that
   copies are not made or distributed for profit or commercial advantage
   and that copies bear this notice and the full citation on the first
   page. Copyrights for third-party components of this work must be
   honored. For all other uses, contact the
   owner\hspace*{.5pt}/author(s).

\noindent Included in the Offline Evaluation for Recommender Systems Workshop (REVEAL'19), collocated with ACM RecSys 2019.

\noindent\textit{REVEAL'19, September 20th, 2019, Copenhagen, Denmark.}

\noindent\copyright  2019 Copyright held by the owner/author(s).
}
} 

\usepackage{balance}

\usepackage[utf8]{inputenc}

\usepackage{calc}

\usepackage{colortbl}
\usepackage{multirow}
\usepackage{hhline}

\usepackage{tagging}

\usepackage{booktabs}

\usepackage{tabularx}
\usepackage{color, colortbl}
\definecolor{Gray}{gray}{0.85}

\usepackage{xcolor}
\usepackage{soul}

\colorlet{soulblue}{cyan!40}
\colorlet{soulred}{red!30}

\usepackage{eucal}
\usepackage{amsmath}         

\usepackage{booktabs} 
\usepackage[inline]{enumitem} 
\usepackage{amsmath}         
\usepackage{eucal}

\usepackage [english]{babel}
\usepackage [autostyle, english = american]{csquotes}
\MakeOuterQuote{"}

\usepackage[english]{babel}
\usepackage{algorithm}
\usepackage[noend]{algpseudocode}

\usepackage[edges]{forest}
\usetikzlibrary{shadows}
\pdfoutput=1

\tagged{comment}{
\usepackage[compact, medium]{titlesec}
\setlength{\textwidth}{1.03\textwidth}

\setlength{\skip\footins}{1.2pc minus 2pt} 
}

\setcopyright{rightsretained}

\acmDOI{10.475/123_4}

\acmISBN{123-4567-24-567/08/06}

\acmConference[REVEAL'19]{REVEAL'19}{September 20th, 2019}{Copenhagen, Denmark}
\acmYear{2019}
\copyrightyear{2019}

\acmArticle{4}
\acmPrice{15.00}

\begin{document}
\title[Towards Sharing Task Environments for Interactive RS Evaluation]{Towards Sharing Task Environments to Support Reproducible Evaluations of Interactive Recommender Systems}


\author{Andrea Barraza-Urbina}
\affiliation{%
\institution{Insight Centre for Data Analytics,\\
Data Science Institute, NUI Galway}
}
\email{andrea.barraza@insight-centre.org}

\author{Mathieu d'Aquin}
\affiliation{%
\institution{Insight Centre for Data Analytics,\\
Data Science Institute, NUI Galway}
}
\email{mathieu.daquin@insight-centre.org}

\renewcommand{\shortauthors}{A. Barraza-Urbina and M. d'Aquin}

\begin{abstract}
Beyond sharing datasets or simulations, we believe the Recommender Systems (RS) community should share \textit{Task Environments}.
In this work, we propose a high-level logical architecture that will help to reason about the core components of a RS Task Environment, identify the differences between Environments, datasets and simulations; and most importantly, understand what needs to be shared about Environments to achieve reproducible experiments. 
The work presents itself as valuable initial groundwork, open to discussion and extensions.
\end{abstract}


\settopmatter{printfolios=true} 
\maketitle

\section{Introduction}

\begin{figure}[t]
\centering
\includegraphics[width=\linewidth]{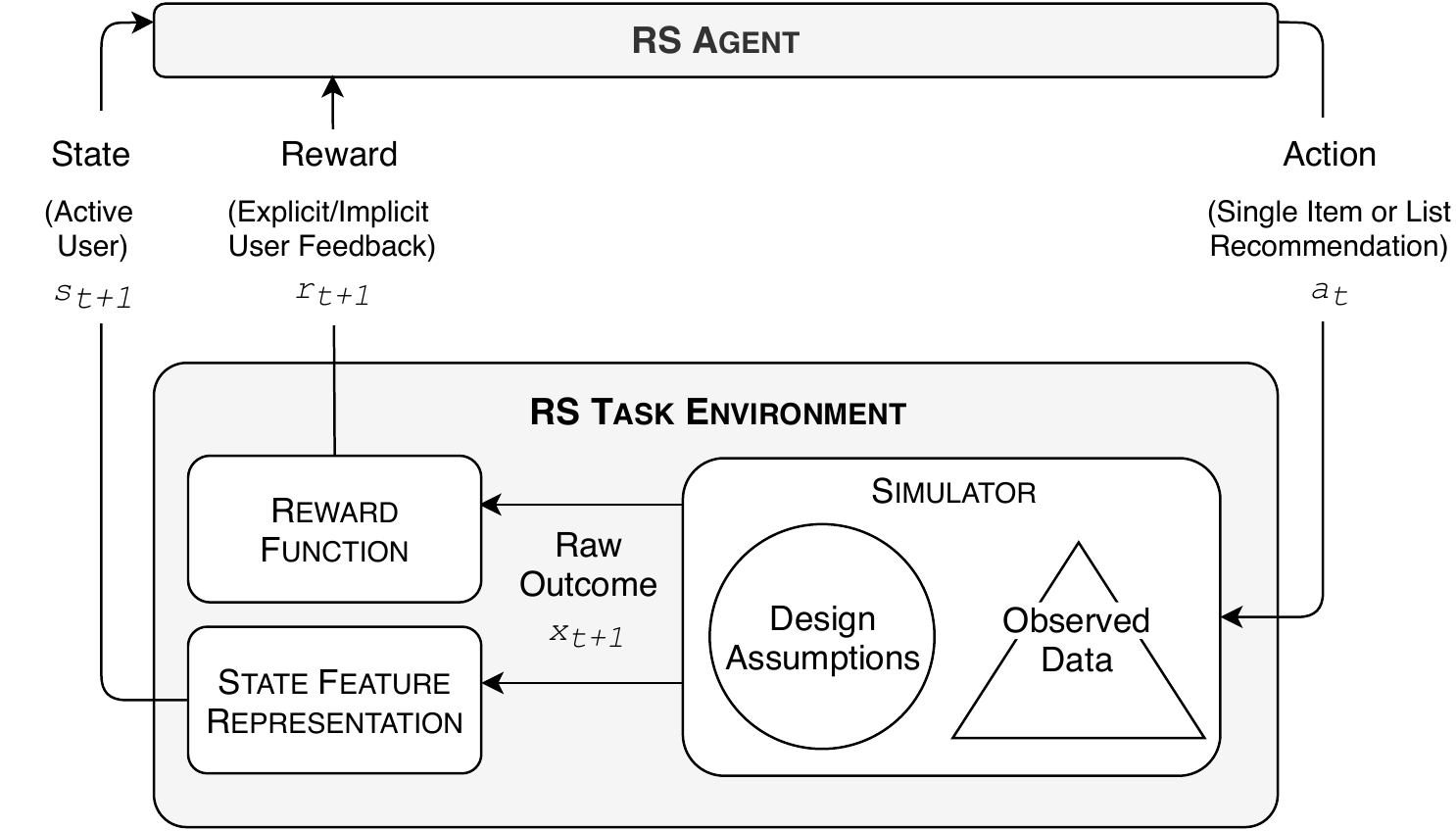}
\caption{
The Interactive Recommender Systems~(RS) problem as Reinforcement Learning~(RL). 
\\
The Figure serves two purposes:
(a)~Present how the RS problem can be mapped to the RL Framework~\cite{rl_book}, showing examples of possible states, rewards and actions in the RS field;
(b)~Focus on the RS Task Environment and present its main components.
}
\label{fig:rl_rs_framework}
\end{figure}

Recommender Systems (RS) help users discover interesting products by means of suggestions.
Conventionally, the RS problem has been formalized as a Supervised Learning task (Batch Learning). 
However, in recent years these foundations have been questioned in light of more complex application settings, such as learning from interactive feedback in fast-paced and dynamic scenarios. 
Recent works, as explained in~\cite{bears}, have proposed that Reinforcement Learning (RL) can be a more appropriate paradigm (Online and Incremental Learning) to frame the ``modern'' and Interactive RS problem.
Under this paradigm, the RS is viewed as a sequential decision-making Agent focused on learning over time how to perform \textit{actions} (e.g., offer item suggestions), in different \textit{states} (e.g., to different users), using interactive feedback in the form of \textit{reward} signals (e.g., user ratings).
Figure~\ref{fig:rl_rs_framework}(a), shows how the RL framework can be used to represent a RS problem (more in~\cite{bears}).

An important component in RL is the Task Environment, which captures the characteristics of the Agent's application domain and expresses the task/goals the Agent is trying to achieve.
Currently, the RS community has mostly focused on creating and sharing datasets and simulations as part of sharing experiments and promoting reproducible research. 
We argue that this is not enough, and that Environments can encapsulate more assumptions and design decisions that are necessarily incurred in RS evaluation design. In short, when sharing only datasets and simulations many evaluation design decisions are typically left to interpretation, which deeply hinders reproducible research.

\section{What is a Task Environment?}
A \textit{RS Task Environment} (or simply \textit{RS Environment}) is the domain or application scenario where the RS Agent will be deployed, and embodies the most relevant characteristics of the considered problem setting.
The Environment is where the Agent operates, and includes everything external to the Agent~\cite{rl_book}.
In its minimal form, the Environment is a function that for any action $a_t$ performed by the Agent over the current Environment state $s_t$, will generate a reward $r_{t+1}$ and the next Environment state $s_{t+1}$, i.e.: $(s_t, a_t) \mapsto (r_{t+1}, s_{t+1})$.

In RL, it is common to describe Environments as a Markov Decision Process (MDP)~\cite{rl_book}. This is useful because independent of the specific implementation details: From the Agent's perspective (in our case the RS) the Environment has a simple interface reduced to the information found in states, actions and rewards.

\section{RS Environment Main Components}
Although we can describe Environments using MDP constructs, in this work, we propose an alternative 
general
framework of components that can help define a principled way to build Environments in the RS field.
Figure~\ref{fig:rl_rs_framework}(b) presents the main components a RS Task Environment should define, which are:


\vspace{0.8em}\noindent
\textbf{Simulator}.\enskip
The \textit{Simulator} imitates the dynamics of the RS application domain (e.g.,~dynamics of users, items and contextual factors). It takes as input the RS Agent's action $a_t$ (e.g.,~recommended items) and outputs a \textit{Raw Outcome} $x_{t+1}$ representing all variables observed after the action is performed over the application. 
In short, it defines the movement from one state of the world to the next given an action.
Note that the Simulator only defines world dynamics (e.g.,~the next active user)
and not the Agent's task/goals.
To build a Simulator, it is common to use \textit{Observed Data}%
\footnote{Alternative names include empirical data, ground-truth data and logged data.}%
(sampled from the real application) augmented with \textit{Design Assumptions} about the world being modeled.
For instance, if the Simulator is based on a static dataset (e.g.,~a user-item matrix), the simulation designer would have to make design assumptions (ideally, based on prior domain knowledge) about the frequency and order in which users interact with the RS.
More examples are presented in the following section.

\vspace{0.8em}\noindent
\textbf{Reward Function}.\enskip
The \textit{Reward Function} uses the Simulator's output $x_{t+1}$ to generate a bounded numeric reward $r_{t+1}$. 
Fundamentally, rewards define the RS goals and are used to motivate the RS Agent to learn an optimal way to act. In fact, typically the RS Agent's goal (objective function) is to maximize a measure of the total reward collected over its interactions with the Environment. 
Thus, designing the Reward Function is one of the most critical tasks of defining an Environment.
In RS, it is common to abstract from $x_{t+1}$ the user's immediate feedback on a recommendation (e.g.,~explicit/implicit rating) to use as the reward signal. Other measures can also be used, e.g.,~number of items sold, abandonment rate, conversion rate, session length or revenue among others.

All in all, it can be particularly challenging to translate high-level complex RS goals to an individual numeric reward signal~\cite{jannach2016recommendations, konstan2018}, and there is much work to be explored in the field of reward engineering for RS. Possible topics are: designing/modeling goals~\cite{littman2017environment,icarte2018using}, reward hacking~\cite{rl_book, amodei2016concrete, ng1999policy}, incorporating artificial curiosity/intrinsic motivation~\cite{singh2009rewards, singh2010intrinsically, barto2013intrinsic, oudeyer2009intrinsic} (e.g.,~by explicitly rewarding the RS for learning new information about users) and inverse RL~\cite{li2017towards}.

\vspace{0.8em}\noindent
\textbf{State Feature Representation}.\enskip
%
The \textit{State Feature Representation} component takes as input the Simulator's output $x_{t+1}$ to generate state $s_{t+1}$.
A state%
\footnote{Also called context in Contextual Bandits, such as~\cite{li2010contextual}.}
encapsulates all the relevant information about the Environment made available to the Agent at a given time step to enable decision making.
An example state in RS, can be represented with a set of \textit{state variables} that include: active user profile (can be an ID), available items (item IDs and possibly features) and information on contextual factors (e.g.,~time, location).
Though RS feature engineering is a fairly mature topic, it is important to fully share how features are identified, selected and represented. Encapsulating this process in a shareable Environment can help support reproducible evaluation.

\section{Why share RS Environments?}

RS address a wide variety of complex problem settings.
Naturally, there are no perfect simulators or complete datasets that can entirely represent all the characteristics of a specific RS problem setting.
As a consequence, evaluation design involves many choices and assumptions, that are not always obvious and are often not made explicit.
RS Task Environments can help encapsulate and share many of the complexities and assumptions behind design choices to support reproducible evaluation.
In principle, by sharing Environments, we could ensure that solution approaches are in fact being tested and compared for the same RS task specification.
%

To illustrate choice complexity, we will offer an example of design choices faced when building a Simulator.
Let us assume we are building a Simulator based on a ground-truth dataset (Observed Data) with data properties that are closely representative of our use case of interest.
When using datasets, we typically face two challenges to build a Simulator: 

\vspace{0.8em}\noindent
\textbf{Data Bias}.\enskip
It is well-known that RS datasets can be biased in multiple ways.
A notable example is selection bias~\cite{schnabel2016recommendations};
as data about users is ordinarily not collected at random 
%
%
(i.e.,~data is missing-not-at-random~\cite{steck2010training}).
For instance, users mostly provide feedback on displayed items, which are naturally influenced by the deployed RS 
(algorithmic bias~\cite{chaney2017algorithmic, adomavicius2014biasing}).
Even if items are presented at random, users can choose to provide, or not, item feedback.
Note that other context biases can influence users, such as
presentation and position bias~\cite{hofmann2016online}.
%


\vspace{0.8em}\noindent
\textbf{Missing Data}.\enskip
Datasets can be missing unobserved confounding variables which can lead to biased results~\cite{hofmann2016online, bottou2013counterfactual}. 
Moreover, it is natural for users to interact with only a small portion of available items, hence datasets really only represent an incomplete picture of user preferences (such as, only bandit feedback~\cite{swaminathan2015batch}).
%
Mainly, evaluation is difficult (sometimes even unfeasible) if actions taken by the new solution approach do not overlap (at least enough) with the actions previously taken by the RS used for data collection~\cite{hofmann2016online}.

\vspace{2.5mm}

In this case, the Design Assumptions circle in Figure~\ref{fig:rl_rs_framework}(b) represents the experiment designer's decisions that aim to: correct bias and complete the data.
There are multiple paths to achieve these goals depending on the dataset. Let's focus on dealing with missing data. We could~\cite{hofmann2016online, jiang2015doubly}:
\begin{enumerate*}[label=(\alph*)]
    \item impute missing rating information assuming data is missing at random, or
    \item compute counterfactual/off-policy estimators, based on the availability of propensity scores or depending on certain assumptions regarding the data collection strategy~\cite{schnabel2016recommendations, li2011unbiased, li2010contextual}
\end{enumerate*}.
If we move forward with our example, we will soon realize that each decision opens the door to further decisions, each with their own set of implications and conditions for success which impact evaluation results in different (sometimes unknown) ways.
Moreover,
the bigger the Design Assumptions circle, the more bias is introduced to the RS Task Environment. Nonetheless, evaluation design assumptions are unavoidable.

This is one of many examples of design decision paths that need to be explicitly shared towards building reproducible experiments. %
%
We believe building RS Task Environments to define and share RS tasks and problem settings can be a reliable solution%
, where design assumptions are necessarily made explicit.

\section{Discussion and Next Steps}
Describing Interactive RS evaluations in a way which is \textit{complete} and \textit{consistent} to be \textit{efficiently reproduced} and \textit{compared}, can quickly turn into an overwhelming task.
In many ways, RS evaluation design is more an art than a science.
\textit{RS Task Environments} can be a way to encapsulate and share many of the complexities and design assumptions necessarily made when
doing RS evaluations.
Recent works such as~\cite{bears,rohde2018recogym} are steps in this direction.
Specifically, this paper introduces a general and modular framework that identifies the main components of a RS Environment, i.e., the shareable components that sufficiently defines a RS problem setting.
The paper also motivates the need to build and share RS Environments to support reproducible evaluation.

We have barely scratched the surface of possibilities, and there are multiple paths for \textit{future research}, such as:
(a)~%
\textit{A Framework to share RS Environments}: BEARS~\cite{bears} presents a possible solution,
(b)~%
\textit{Defining types of RS Environment:}
Individual components can be built and combined to create multiple types of RS Environments.
%
Naturally, a Simulator that is not based on Observed Data (i.e.,~big Design Assumptions circle in Figure~\ref{fig:rl_rs_framework}(b), aka higher model bias) would have different properties compared to a Simulator that only uses Observed Data (e.g.,~an online experiment requiring less assumptions about how data is generated, aka lower bias).
Also, different Reward Function and State Feature Representation components can be designed for the same Simulator component.
Though there seem to be infinite possibilities, we expect that these types of Environments share some common properties and that it will be valuable to create a taxonomy of RS Environments, for instance, to understand which types of Environments are comparable,
(c)~%
\textit{Assessing RS Environment Quality}:
We cannot build perfect Environments, thus it is important to analyze Environments to understand in which ways they are imperfect and which problem settings they represent best (e.g.,~to identify benchmark RS Environments for different domains and/or tasks).
Note that
different types of Environments lead to different bias-variance trade-offs which need to be understood.
%
Overall, we hope that the introduced model helps structure a discussion around the meaning of Task Environments for RS, their role in evaluation design and encourages new research in this area.








\begin{acks}
This publication has emanated from research conducted with the financial support of Science Foundation Ireland (SFI) under Grant Number SFI/12/RC/2289\_P2, cofunded by the European Regional Development Fund.
\end{acks}

\balance
\bibliographystyle{ACM-Reference-Format}
\bibliography{references} 

\end{document}